\documentclass[aps,twocolumn,groupedaddress,showpacs,floatfix]{revtex4}

\usepackage{graphicx}
\usepackage{amsmath}

\begin{document}

\title{Agile low phase noise radio-frequency sine wave generator applied to experiments on ultracold atoms}
\author{O.~Morizot}
\author{J.~de Lapeyre de Bellair}
\author{F.~Wiotte}
\author{O.~Lopez}
\author{P.-E.~Pottie}
\author{H.~Perrin}\email{helene.perrin@galilee.univ-paris13.fr}

\affiliation{Laboratoire de physique des lasers, Institut Galil\'ee, Universit\'e Paris 13 and CNRS,
Avenue J.-B. Cl\'ement, F-93430 Villetaneuse, France }

\date{\today}

\begin{abstract} We report on the frequency performance of a low cost ($\sim$500~\$) radio-frequency sine wave
generator, using direct digital synthesis (DDS) and a field-programmable gate array (FPGA). The output frequency of the
device may be changed dynamically to any arbitrary value ranging from DC to 10~MHz without any phase slip. Sampling
effects are substantially reduced by a high sample rate, up to 1~MHz, and by a large memory length, more than $2\times
10^5$ samples. By  using a low noise external oscillator to clock the DDS, we demonstrate a phase noise as low as
that of the master clock, that is at the level of $-113$~dB.rad$^2/$Hz at 1~Hz from the carrier for an output
frequency of 3.75~MHz. The device is successfully used to confine an ultracold atomic cloud of rubidium~87 in a
RF-based trap, and there is no extra heating from the RF source.

\end{abstract}


\maketitle

\section{Introduction}
Radio-frequency (RF) fields are used in cold atom experiments for different purposes: for instance, evaporative cooling
performed in a magnetic trap relies on RF field coupling between the different atomic magnetic
states~\cite{Hess1986,Hess1988}. This technique led to the first observation of Bose-Einstein condensation
(BEC)~\cite{Cornell2002,Ketterle2002}. Also, RF pulses are used for dissociating ultracold molecules produced from
ultracold gases through Feshbach resonances~\cite{Regal2003}. More recently, RF fields have been used together with
static magnetic fields for trapping utracold atoms at a temperature of a few $\mu$K in unusual
geometries~\cite{Zobay2001,Colombe2004}. There is a growing interest for these ``RF-based traps'' among atomic
physicists, for creating double well traps on atom chips~\cite{Schumm2005,Extavour2006,Jo2007} or proposing new kinds
of confining potentials~\cite{Lesanovsky2006,Courteille2006,Morizot2006}. In both cases, a single frequency RF signal
must be frequency swept over some range, often larger than the initial frequency, following a precise time function
lasting several seconds. Typically the RF frequency is varied between 1~MHz and a few tens of MHz in 0.1~s to 10~s in
the ramping stage, and held at the final frequency for seconds in the plateau stage. For cooling purposes, commercial
RF generators fit physicists' needs reasonably well, even if a better resolution in arbitrary frequency ramps would be
appreciated. However, in the case of RF-based trapping, the requirements are stronger. The main difference between
these two situations is as follows: in evaporative cooling the cold atomic sample is located away from the region of
efficient coupling, whereas in the RF-based trapping scheme the atoms sit exactly at the point where the RF field has
the largest effect. The quality of the RF source is then much more important than for evaporative cooling. In fact, the
cloud position is directly related to the value of the RF field frequency, and the trap restoring force, or
equivalently the oscillation frequency $\nu_t$ in the harmonic approximation, is linked to the RF amplitude. As a
result, any amplitude noise, frequency noise or phase noise of the RF signal during the ramp or the plateau leads to a
heating of the cold atomic cloud. This motivated the construction of a synthesizer fitting our requirements.

This paper is organized as follows. In section~\ref{Sec:Theory} we give explicit expressions for the heating of the
cold atom sample for frequency and amplitude noise in the case of RF-based trapping. In
section~\ref{Sec:DeviceDescription}, we describe our RF synthesizer. Finally, section~\ref{Sec:Results} is devoted to
experimental results on its performance and comparison between the different RF sources tested on the BEC experiment.

\section{Requirements on the RF source for RF-based trapping}
\label{Sec:Theory}
In this section, we will focus on the RF-dressed trap that we experimentally produce in the
laboratory~\cite{Colombe2004}. The extension of the main conclusions to other RF-dressed trap geometries is
straightforward.

The trap confines the atoms in all three space dimensions. The trapping force arises from the interaction between the linearly polarized
RF field $B(t) = B_{\rm RF} \cos(2\pi \nu_{\rm RF} t)$ and the atoms in the presence of an inhomogeneous magnetic field. This interaction results in a transverse
confinement of the atoms to the surface of an ellipsoid. The atoms are free to move along the confining surface,
resulting in a kind of ``bubble trap''~\cite{Zobay2001}. Due to gravity, however, the atoms are concentrated at the
bottom of the ellipsoid. Their motion is pendulum-like in the horizontal directions, and imposed by the RF interaction
along the vertical $z$ axis. This last direction is thus the most sensitive to the RF field properties (frequency
$\nu_{\rm RF}$, amplitude $B_{\rm RF}$) and we will concentrate on the vertical motion in the following. Along this
direction, heating or atomic losses may arise from frequency or amplitude noise, phase hops or sudden frequency hops
during the RF ramp.

\subsection{Dipolar excitation heating}

Very generally, for atoms in a one dimensional harmonic trap with a trapping frequency $\nu_z$,
any effect producing a jitter in the trap position $z$ results in linear heating through
dipolar excitation. The average energy of the cold atomic cloud $E$ increases linearly
as~\cite{Gehm1998}:
\begin{equation}
\dot{E}=\frac{1}{4} M \omega_z^4 \, S_{z}(\nu_z) \label{eqn:dipolar_heating}
\end{equation}
where $\omega_z = 2 \pi \nu_z$, $M$ is the atomic mass and $S_{z}$ is the one-sided Power Spectral Density (PSD) of the
position fluctuations $\delta z$, defined as the Fourier transform of the time correlation function~\cite{Gehm1998}
\begin{equation}
S_z(\nu) = 4 \int_0^{\infty} \! \! d\tau \, \cos(2 \pi \nu \tau) \langle \delta z(t) \, \delta
z(t+ \tau) \rangle. \label{eqn:Sz}
\end{equation}
The time variations of energy, $E$, and temperature, $T$, are related by $\dot{T}=\dot{E}/3k_B$. The factor 3 arises
because only one degree of freedom is responsible for the temperature increase, as is the case in our atom trap. The
vertical trap position $z$ is linked to the RF frequency $\nu_{\rm RF}$ by $z = Z(\nu_{\rm RF})$ such that $S_{z}$ is
directly proportional to $S_{y}$, the PSD of relative frequency noise of the RF source, through:
\begin{equation}
S_z(\nu) = \left(\nu_{\rm RF} \frac{d Z}{d \nu_{\rm RF}} \right)^2 S_y(\nu).
\label{eqn:Sz_to_Sy}
\end{equation}
The function $Z$ depends on the geometry of the static magnetic field. In a quadrupolar field, for instance, $Z$ is
linear with $\nu_{\rm RF}$ and its derivative is simply a constant. From Eqs.~(\ref{eqn:dipolar_heating}) and
(\ref{eqn:Sz_to_Sy}), we infer that the linear heating rate is proportional to $S_{y}(\nu_z)$.

To fix orders of magnitude, within the static magnetic field of our Ioffe-Pritchard trap~\cite{Colombe2004}, $\nu_z$
may be adjusted between 600 and 1500~Hz and the typical temperature of the cold rubidium 87 atoms ranges from 0.5 to
5~$\mu$K. To maintain a temperature below condensation threshold for a few seconds, a linear temperature increase below
0.1~$\mu$K/s is necessary. This rate corresponds to $\sqrt{S_{z}(\nu_z)} = 0.27$~nm/$\sqrt{\rm Hz}$ for an intermediate
trap frequency of 1000~Hz and $\nu_{RF} = 3$~MHz, which in turn corresponds to a one-sided PSD of relative frequency
fluctuations of the RF source $S_y(\nu_z) = -118$~dB/Hz.

\subsection{Parametric heating}
Fluctuations of the RF field amplitude $B_{\rm RF}$ are responsible for parametric heating in the vertical direction.
The trapping frequency $\nu_z$ is inversely proportional to $\sqrt{B_{\rm RF}}$~\cite{Zobay2001}:
\begin{equation}
\nu_z = \left(\frac{dZ}{d\nu_{\rm RF}}\right)^{-1}\sqrt{\frac{2F\hbar}{M\gamma B_{\rm RF}}} .
\end{equation}
Here, $\gamma$ is the gyromagnetic ratio of the atom and $F$ is the total atomic spin ($F=2$ for rubidium 87 in its upper hyperfine state). The atoms are assumed to be polarized in their extreme $m_F=F$ substate. The cloud temperature increases exponentially
due to amplitude noise with a rate $\Gamma$, where
\begin{equation}
\Gamma = \pi^2 \nu_z^2 S_a(2 \nu_z)
\end{equation}
and $S_a$ is the PSD of the relative RF amplitude noise~\cite{Gehm1998}. In order to perform experiments with the BEC
within a time scale of a few seconds, $\Gamma$ should not exceed $10^{-2}$~s$^{-1}$. Again, for a typical oscillation
frequency of 1000~Hz, this corresponds to $S_a < -90$~dB/Hz. This requirement is rather easy to match and does not
limit the choice of the RF source, as -110~dB/Hz is commonly reached. However, particular care must be taken in the
choice and installation of the RF amplifier usually used after the source.

\subsection{Phase hops}
Controlling the phase of the RF source is not a crucial point for evaporative cooling, but is an issue in the case of
RF-based traps, where it is associated with trap losses. In the latter situation, the atomic spin follows an effective
magnetic field oscillating at the RF frequency. A phase hop results in a sudden flip of this effective field, the
atomic spin being then misaligned with the new direction of the field. Some of the atoms end up with a spin oriented
incorrectly and escape the trap.

For this reason, phase hops should be avoided. This is difficult to achieve with an analog synthesizer over a wide
frequency sweep. By contrast, Direct Digital Synthesis (DDS) technology is well adapted to this
requirement\,\cite{DDSBook}.

\subsection{Frequency steps}
\label{Sec:FrequencySteps} The drawback of DDS technology is that, although the phase is continuous, the frequency is
increased by $N$ successive discrete steps $\delta\nu$. A sudden change in the RF frequency also results in atomic
losses, through the same mechanism as for phase hops. The effective magnetic field rotates, at most, by the small angle
$\delta\theta = 2\pi\,\delta\nu/(\gamma B_{\rm RF}/2)$~\cite{note1}. For a linear ramp
over a frequency range $\Delta\nu = N \delta\nu$, the  fraction of atoms remaining after the full ramp is of order $(1
- F \delta\theta^2/2)^N$. Given
the expression for $\delta\theta$, this reads:
\begin{equation}\label{Eq:FreqStepCriteria}
\left(1 - \frac{F}{2} \left(\frac{4\pi\,\Delta\nu}{N\gamma B_{\rm RF}}\right)^2\right)^N \simeq 1 -
\frac{F}{2N} \left(\frac{4\pi\,\Delta\nu}{\gamma B_{\rm RF}}\right)^2 \, .
\end{equation}
Thus, for the remaining fraction to be larger than 95\%, the number of frequency steps should be larger than $10 F
(4\pi\,\Delta\nu/\gamma B_{\rm RF})^2$. For example, for a 2~MHz ramp with a typical RF amplitude of 200~mG, $N$ should be larger than 16,000.

In addition to this loss effect, a sudden change in the RF frequency results in a sudden shift of the position of the
RF-dressed trap.  This may cause dipolar heating of the atoms, especially if this frequency change occurs every trap
period. The frequency steps should thus be as small as possible, a few tens Hz to a hundred Hz typically.

\section{Device description}
\label{Sec:DeviceDescription} Our experiment has the following requirements. First, during the ramp the gap
between two successive frequencies must fulfil the criterion discussed in section~\ref{Sec:FrequencySteps}. Second,  adiabaticity criteria require a controlled, optimized ramp. Third, the ramp duration
should be tunable from one experiment to the other on a time scale ranging from 50~ms to 10~s. Finally, frequency and
amplitude noise must be small enough, as discussed in previous section.

Given the amplitude of the frequency sweep we need to perform in our experiment, DDS technology appears to be an ideal
solution. We previously used a commercial DDS-based RF generator, the Stanford DS-345. Its memory length is limited to
1,500 frequency points for each waveform with an adjustable step duration of 40~Msample/s/$N$, with $N$=1 to
$2^{34}-1$. The major inconvenience of this device is that it is unable to hold the final frequency at the end of the
ramp. Instead, the frequency sweep is looped indefinitely. It forced us to sacrifice either frequency resolution
during the ramp or duration of the plateau. To benefit from both a low noise RF spectrum during the ramp and a
very small frequency step, and to improve the possibilities of the RF source, we designed a digital RF synthesizer with
a $>200,000$ memory length and great agility, fitted to our experimental requirements.

The main features of the RF synthesizer are as follows. It is able to generate 262,144 sine waves in a row in the radio
frequency band (DC - 10~MHz), owing to its 1~M-byte fast asynchronous Static Random Access Memory (SRAM). Each
frequency is an integer chosen by the user. A key feature of the device is a variable sample frequency over
the sequence, as the duration of each generated frequency can be tuned from 1~$\mu$s to 1 hour.

\begin{figure}[ht]
    \begin{center}
        \includegraphics[width=0.95\columnwidth]{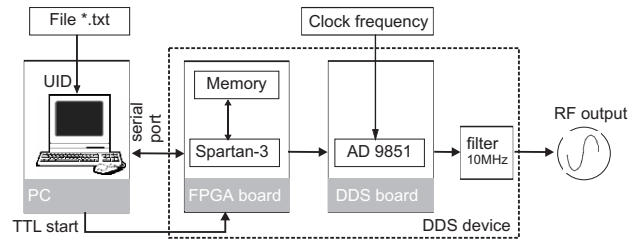}
    \end{center}
        \caption{Layout of the system.
        \label{fig:Layout}}
\end{figure}

The general architecture of the device is sketched in Fig.~\ref{fig:Layout}. It is made up of one evaluation kit DDS
board, and a ``starter kit'' Field Programmable Gate Array (FPGA) board. The device is managed by a Personal Computer
(PC). The DDS is clocked by an ultra-stable external reference signal. The output of the DDS is a sine wave, filtered
through a 10~MHz low-pass filter. The frequency ramp synthesis starts when a TTL signal is sent to the device.

The DDS board combines digital parameters and an analog reference clock frequency to generate a sine
wave~\cite{DDSBook}. The heart of the DDS board is a digitally programmable device using DDS technology, the AD9851. It
has a 32-bit phase accumulator, a 14-bit digital phase-to-amplitude converter and a 10-bit Digital-to-Analog Converter
(DAC). Its maximum clock frequency is 180~MHz, and its maximum output frequency is 70~MHz. The phase, relative to the
clock signal, is encoded in 5 bits, and is adjustable to any value from 0 to $2\pi$. This results in a rather poor phase
resolution of 196~mrad.

The FPGA board manages the 1~Mb memory, the time settings and the input/output of the device through serial port. The
FPGA is a Xilinx spartan-3 XC3S200, providing $200,000$ logic gates. These logic gates are designed with
VHDL~\cite{note2}. A Universal Asynchronous Receiver
Transmitter(UART) and a Picoblaze microcontroller are loaded into the FPGA, in order to communicate through serial port
to the PC and to load the on-board memory. We wrote our own VHDL scripts to manage the DDS board and the FPGA in-board
memory. The FPGA board is clocked internally at 50~MHz. The output sample rate of the device depends on the number of
clock cycles $n_{\rm cycle}$ between frequency data transferred to the DDS board. We set $n_{\rm cycle}=50$
so that $n_{\rm cycle}/50~{\rm MHz}=1~\mu{\rm s}$, large enough thus ensuring safe operation of the frequency data.

Software was developed in C with CVI Labwindows in order to configure the device. The user writes a plain text file
ordering all the frequencies of the desired frequency ramp. The set of frequencies is separated into 10 groups of
adjustable length, with a given sample rate for each. The group lengths and the corresponding sample rate are each translated
by the software into 4 bytes. In addition, each frequency in a given group, an integer written as a decimal number, is
translated into 4 bytes (32 bits). The software sends these bytes by serial port to the FPGA board.

The clock frequency $f_{c}$ and its phase noise level are the key points for setting the frequency performance of the
device. The lower the phase noise of the clock signal, the lower is the {\it minimum} phase noise of the output
frequency of the device (see section \ref{SubSec:FreqPerf}).

The clock signal used for the experiment, see next section, is the 10~MHz clock signal from an ultra-stable Oven
Controlled Crystal Oscillator (OCXO) BVA-8600. Its phase noise PSD is $-115$~dB.rad$^{2}$/Hz at 1 Hz. As this clock
frequency is very close to the desired maximum output frequency ($\simeq$~10~MHz, see next section), and to fulfil
Shannon's theorem~\cite{Nyquist1928,Shannon1949}, we generate a higher clock frequency by using the internal
frequency clock multiplier, at $\times 6$, of the DDS board.

\section{Results}
\label{Sec:Results}

The device presented in section~\ref{Sec:DeviceDescription} was first tested for its frequency stability performance, as described in section~\ref{SubSec:FreqPerf} and summarized in Table~\ref{Tab:perfo}.
It was then integrated into a Bose-Einstein condensation experiment, see section~\ref{2Dtrap}.

\subsection{Device frequency performance}
\label{SubSec:FreqPerf}

\subsubsection{Quantization error, Phase Accumulator truncation and ``{\it magic}'' frequencies} By construction,
digitization yields to inaccuracies in frequency synthesis. The output frequency of 32-bit resolution DDS is given by
\begin{equation}\label{Eq:Sampling}
f_{RF}=f_{c}\times \frac{w}{2^{32}}
\end{equation}
where $w$ is a binary 32-bit tuning word. The output frequency can thus  differ slightly from the desired frequency. As
$f_{c}$ = 60~MHz, the maximum digitization error $\delta f$ is $6\times10^{7}$/$2^{32}$ = 0.014~Hz.

As our software only takes integer frequencies as input, a given frequency $f_{RF}$ will be synthesized
without sampling error if it may be written exactly as an integer in the form given at Eq.~(\ref{Eq:Sampling}). This
condition is written as:
\begin{equation}
w=n\times 2^{32-p}
\end{equation}
where $n$ is a positive integer and $p$ is the power of 2 in the prime factorization of the clock frequency $f_{c}$. In
our case, $f_{c}=2^{8}\times 3\times 5^{7}$~Hz and $p=8$ so that every frequency verifying
\begin{equation}\label{Eq:Condition1}
f_{RF}=n\times 234375~\mathrm{Hz}
\end{equation}
will yield to no digitization error. $n$ should be less than $2^p/3$ for the desired frequency to be in the synthesizer
range ($f_c/3$).

In addition, when the AD9851 converts the calculated phase to an effective output amplitude, only the first most
significant 14 bits are used, even though the AD9851 is a 32-bit synthesizer, in order to handle practicable number of
entries in a lookup table. Truncating the phase results in errors in amplitude that are periodic in the time domain.
These errors will be seen as spurs in the frequency domain. However, for particular frequencies which are exactly
encoded by the first 14 bits (The last 18 bits are 0.), the phase is not truncated at all, yielding no spurious effects
and the best PSD of phase noise. This occurs for every frequency satisfying
\begin{equation}
\label{Eq:Condition2} f_{RF}=n\times f_{c}/2^{14}
\end{equation}
with $n$ a positive integer. As $f_{c}$ = 60~MHz, we have $f_{RF}=n\times 3662.109375$~Hz. The most stringent condition
being the first one, we will denote the frequencies satisfying Eq.~(\ref{Eq:Condition1}) as ``magic frequencies''. In
order to illustrate the difference between a ``magic frequency'' and another one, we performed a set of noise PSD
measurements for two frequencies: a first set for $f_{RF}=3$~MHz which is not a ``magic'' frequency, and a second set for
$f_{RF}=3.75$~MHz which is a ``magic'' frequency.

\begin{table}
\caption{Performance of the device with $f_{c}= 60$~MHz. The relative frequency noise is computed from the phase noise
data, and given for 3.75~MHz (``magic'' frequency) and 3~MHz (larger noise value).\label{Tab:perfo}}
 \begin{ruledtabular}
     \begin{tabular}{lccr}
     Parameters & Min. & Max. & Units\\
     \hline
     Dynamic & 0 & 10 & MHz\\
     Line-width & - & 30 & mHz\\
     Digitalization error & 0 & 14 & mHz\\
     Sample rate & adjustable &1 & MHz \\
     Memory length & 1 & 262,144& pts \\
     Phase noise@1Hz & -113 & -78  & dB.rad$^{2}$/Hz \\
     Rel. freq. noise@1Hz & -244 & -207 & dB/Hz \\
     Rel. ampl. noise@1Hz & - & -120 & dB/Hz \\
      \end{tabular}
 \end{ruledtabular}
\end{table}

\subsubsection{Spectral density of noise}

\begin{figure}[b]
 \begin{center}
 \includegraphics[width=0.95\columnwidth]{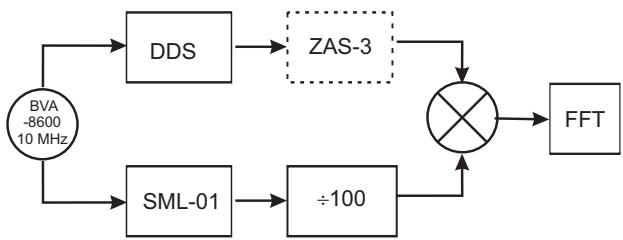}
 \end{center}
\caption{Noise spectrum measurement bench. The gain in the mixer was fitted to the output
signal, depending on the phase between the two beating signals.\label{fig:SketchBeatMeas}}
\end{figure}

\begin{figure}[t]
 \begin{center}
 \includegraphics[width=1.05\columnwidth]{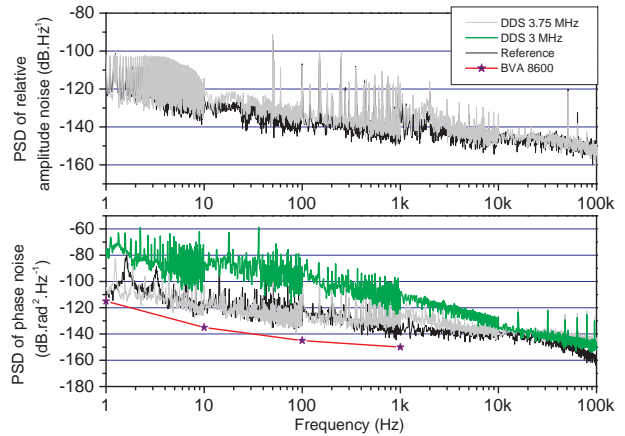}
 \end{center}
\caption{PSD of relative amplitude noise, upper plot, and phase noise,lower plot. Averaging $=100$. FFT's
sampling frequency $=256$~kHz. The low pass filter cutting frequency of the mixer was set to 200~kHz. The reference
curve corresponds to the beat note of two identical R\&S synthesizers. Plotted for comparison is
the phase noise specifications of the BVA 8600 quartz oscillator, as given by the manufacturer. \label{fig:spectra} }
\end{figure}

We recorded the spectral density of noise of our synthesizer at a given frequency $f_{RF}$ by FFT analysis of the beat
note at 0~Hz with a second synthesized reference signal. The measurement bench is sketched in
Fig.~\ref{fig:SketchBeatMeas}. In order to generate a tunable reference signal in the RF range we used an analog
synthesizer, Rhode \& Schwartz SML-01 (R\&S), for synthesizing a signal at a high frequency, and then divided by 100 to
give $f_{RF}$ for subsequent mixing. The beat note was recorded and analyzed by a digital FFT analyzer HP 3562A
sampling at 256~kHz. The R\&S, as all the measurement devices, was clocked at 10~MHz by the ultra-stable BVA-8600
quartz oscillator. All the plots shown on Fig.~\ref{fig:spectra} are raw spectra of the beat note. The reference signal
(R\&S) itself was characterized by making a beat note with a second identical R\&S based synthesis. This corresponds to
the line labelled as ``Reference'' in Fig.~\ref{fig:spectra}.

By tuning the phase difference $\phi$ between the RF signal and the reference signal to $\pi/2$, we recorded phase
noise. At $f_{RF}=3$~MHz, the PSD of phase noise is $-78$~dB.rad$^2.{\rm Hz}^{-1}$ at 1~Hz, which corresponds to a PSD
of relative frequency noise of $-207$~dB.Hz$^{-1}$. From the data reported on Fig.~\ref{fig:spectra}, and assuming a
Lorentzian lineshape for the beat note, the linewidth $\delta f$ of the RF signal, given by\:
\begin{equation}
\delta f = \pi \nu_{RF}^{2}\frac{1}{\Delta f }\int_{\Delta f} \! \! df \, S_{y}(f) \, ,
\end{equation}
is as small as 30~mHz over a bandwidth $\Delta f=100$~kHz for an output frequency of 3~MHz~\cite{Hal1971}. We found
similar results for output frequencies from 1 to 5~MHz. At a ``magic'' frequency, as for example at 3.75~MHz, where
truncation effects cancel out, the results are even better, with a PSD of phase noise as low as $-113$~dB.rad$^2.{\rm
Hz}^{-1}$ at 1~Hz, only 2~dB higher than the phase noise of the BVA 8600. The observed value corresponds to the
ultimate phase noise of the DDS chip specified by the manufacturer. The relative frequency noise is then
$-244$~dB.Hz$^{-1}$.

The frequency noise performance is naturally linked to the quality of the master oscillator. To illustrate this fact,
the same measurements were repeated with the improved OCXO of a DS-345, the ERC EROS-750-SBR-4, as master oscillator.
No significant change in the frequency performances were noticed at non magic frequencies. At a magic frequency, the
PSD of phase noise increased to $-100$~dB.rad$^2$/Hz, which is consistent with the phase noise specifications of this
quartz.

The PSD of relative amplitude noise $S_{a}$ was recorded with the same measurement bench, by tuning the phase $\phi$ to
0. The reference signal was also delivered by the R\&S. The recorded spectrum is very close to the ``Reference'' line
itself at the level of $-120~\mathrm{dB.Hz}^{-1}$ at 1~Hz, close to the input noise of the FFT analyzer. Note that for
relevant excitation frequencies $\nu=2\nu_{z}$ (larger than 1.2~kHz), the PSD of amplitude noise is lower than
$-130$~dB.

In practice, we use in our BEC experiment a programmable RF attenuator Minicircuit ZAS-3, driven by an analog output
channel of a National Instrument PC card PCI-6713, in order to vary the RF amplitude sent to the RF coil. We repeated
the measurements described above with this attenuator and found an increase of $S_{a}$, starting at
$-115~\mathrm{dB.Hz}^{-1}$ at 1~Hz, which corresponds to a noise figure of the programmable attenuator of 5.1~dB. At
the output of the attenuator, the RF signal is amplified by a class-A amplifier Kalmus 505 F. Its gain is 37~dB and
its noise figure is typically +10~dB according to the manufacturer specifications.

\subsection{Integration into the experiment}
\label{2Dtrap}
In this section, we present the main experimental results concerning the heating rate of the atomic sample.

\begin{figure}[t]
 \begin{center}
 \includegraphics[width=0.92\columnwidth]{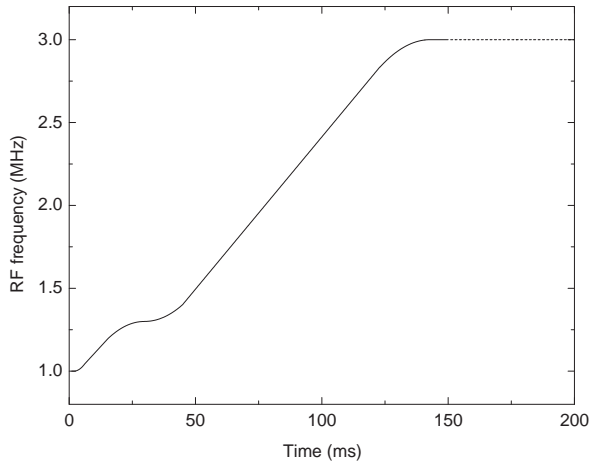}
 \end{center}
\caption{Typical shape of a radio-frequency ramp applied to the ultracold atomic sample. In the present
example $f_{RF}$ is increased from 1 to 3~MHz within 150~ms. At the end of this ramp, the RF frequency is maintained at
its final value for some holding time in the RF-based trap, dashed line.}
\label{fig:ramp}
\end{figure}

As described in section~\ref{Sec:Theory}, the RF signal is used for producing the bubble-like trap, where ultracold
rubidium atoms are accumulated at the bottom of the surface. The trap is very anisotropic with stronger
confinement in the vertical direction~\cite{Colombe2004}. For the RF-dressed trap to be efficiently loaded from the
standard Ioffe-Pritchard static magnetic trap, described in~\cite{Colombe2003}, the RF frequency $f_{RF}$ has to be
ramped up from 1~MHz to a final fixed frequency $f_{RF}^{end}$ ranging from 2 to 10~MHz. The static magnetic field,
necessary both for magnetic trapping and RF-induced trapping, is always present. A typical ramp is shown on
Fig.~\ref{fig:ramp}. The frequency is ramped more slowly around 1.3~MHz, corresponding to the resonant frequency at the
center of the magnetic trap where adiabaticity of spin rotation is more difficult to obtain. At the end of this ramp,
which may last between 75~ms and 500~ms, the RF frequency is held between 0.1 and 10~s for
testing the lifetime and heating rate of the atoms in the RF-based trap.

The RF signal is amplified by 37~dB with a single stage amplifier, and the RF field is produced by a
small circular antenna. The RF field is linearly polarized and its amplitude $B_{\rm RF}$ may be adjusted between 70 and 700~mG. We
record the atomic temperature after the RF ramp as a function of time while the atoms are confined in the RF-based
trap. The temperature is deduced from the cloud size along $z$ after a 7~ms of ballistic expansion. The same measurement
was repeated with different RF sources.

\begin{figure}[t]
 \begin{center}
\includegraphics[width=\columnwidth]{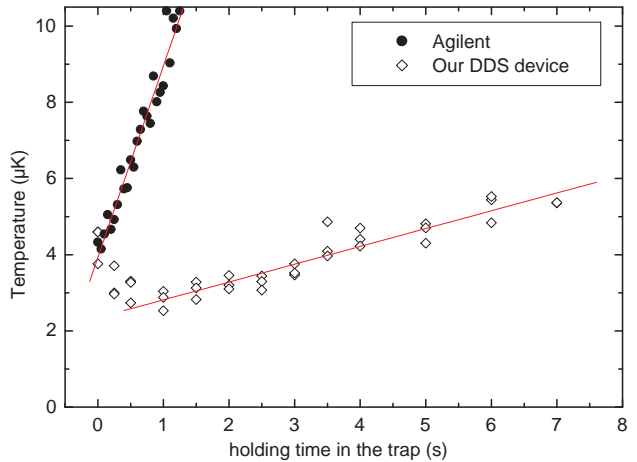}
 \end{center}
\caption{Comparison of heating of the atomic cloud in the bubble trap: Agilent 33250A synthesizer, full circles, or present device, open diamonds, is used for creating the RF ramp and the
final radio-frequency $f_{RF}^{end}$. We observe a heating rate of 5.0~$\mu$K/s in the first case and 0.47~$\mu$K/s the
second one, as given by a linear fit, full lines.} \label{fig:comparaisonHeating}
\end{figure}

First, we used a Agilent 33250A synthesizer with a RF amplitude larger than 500~mG for both the frequency ramp and the
final holding frequency. Such RF analog synthesizers operated at fixed frequency exhibit very good relative
frequency noise in most cases, typically at the $-180$~dB/Hz level or better. However, as mentioned by Colombe \emph{et
al.}~\cite{Colombe2004} and confirmed by White \emph{et al.}~\cite{White2006}, the relative frequency noise increases
by a few decades if the output frequency is driven with an external analog voltage. The frequency was indeed tuned
through an external voltage control provided by the PC analog board (NI 6713),  such that the modulation depth was
$\pm$1~MHz on a central frequency of 2~MHz. We obtained both a short lifetime, typically 400~ms at $1/e$, and a strong
linear heating, as shown on Fig.~\ref{fig:comparaisonHeating} full circles. The heating rate is measured to be
$5.0~\mu$K/s. This rate, given the RF amplitude, corresponds to a relative frequency noise of $S_y=-100$~dB/Hz at the
trap frequency of 600~Hz. This noise is quite high and is related to voltage noise on the external frequency control of
the synthesizer. This effect is strong in our case as the frequency is varied with a large modulation depth ($\Delta
f/f = 1$).

\begin{figure}[t]
 \begin{center}
\includegraphics[width=\columnwidth]{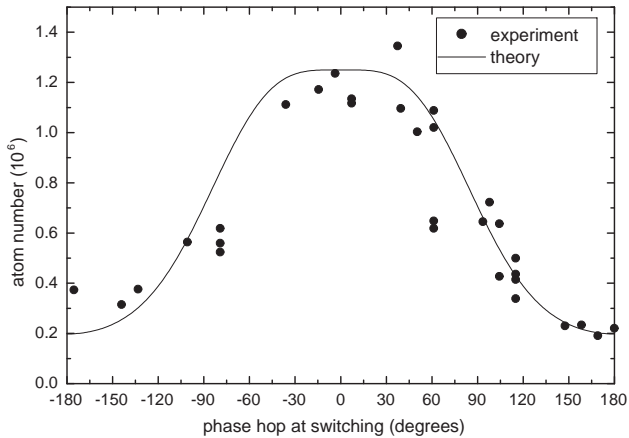}
 \end{center}
\caption{Number of atoms remaining after switching between the two synthesizers as a function of the phase hop.
Experimental data, full circles, are compared to a calculation, full line, assuming a RF amplitude of 470~mG and a
temperature of $4~\mu$K.}
\label{fig:phasehop}
\end{figure}

We also tested a two step scheme, with a first ramp performed by a Stanford DS-345 DDS (1,500 frequency points),
followed by the R\&S maintained at fixed frequency for the full holding time. This scheme allows one to benefit from
the excellent frequency stability of the second device used at fixed frequency. With this setup, heating during the
plateau stage seemed to be completely suppressed~\cite{Colombe2004}. However, a large dispersion was observed in the atom number
data after the switching, which prevented us from characterizing the heating very precisely. This dispersion
is due to a random phase hop at the switching time between the two synthesizers, resulting in atomic losses. We studied
the effect of the random phase hop on the atom number by recording, for each experiment, the phase difference at the
switching time with a control oscilloscope. The results are presented on Fig.~\ref{fig:phasehop}, full circles. For the maximum phase hop, $\pi$,
$80\%$ of the atoms are lost. This figure depends on the atomic temperature, the losses
being higher at lower temperature, and is well reproduced by theory, as shown on Fig.~\ref{fig:phasehop}, black line.
The theoretical curve is calculated for an RF amplitude of 470~mG by averaging the loss probability over the positions of the atoms, as deduced from
a thermal distribution at a temperature of $4~\mu$K. The fact that the trap is able to
hold two of the five spin components of the $F=2$ hyperfine state is taken into account.

Finally, we performed the heating rate and lifetime measurements with our DDS device and an amplitude of 70~mG. We
found an increased lifetime, up to 9.6~s. During the first second, the vertical ``kinetic temperature'', deduced from the
vertical cloud size after time of flight, decreases due to thermalization with the horizontal degrees of freedom which
initially have a lower energy due to the loading procedure: Fig.~\ref{fig:comparaisonHeating}, open diamonds.
After 1 second, a linear heating rate of 0.47~$\mu$K$.s^{-1}$ is clearly observable. No exponential parametric
heating is measurable. We believe that the remaining linear heating is due to residual excitation by scattered light,
which was not present in the previous case. The same heating rate was indeed observed within the static magnetic trap,
when the RF source was off. Heating directly linked to the RF source is not observable.

\section{Conclusion}
We built a low cost DDS RF generator, based on a DDS chip and its evaluation kit. It is extremely agile since we have
the ability of using up to 10 different time steps during the sequence, allowing an efficient use of the FPGA on-board
memory. In contrast to Stanford DS-345 RF synthesizers, the frequency ramp is not repeated at the end of the sequence.
The last frequency is kept until a new TTL pulse is proceeded to end the sequence. Its large memory length, up to
262,144 frequency samples, allows very small frequency steps, thus strongly reducing the atomic losses in our RF-dressed trap
experiment. The DDS technology ensures that the phase is continuous all over the frequency ramp, which is essential for
RF trapping applications.

Apart from these high resolution features, the device presents excellent spectral properties. The power spectral
density of phase noise is as low as $-113$~dB.rad$^2$.Hz$^{-1}$ at well chosen ``magic frequencies'', and
in any case remains below than $-78$~dB.rad$^2$.Hz$^{-1}$ at worst, dominated by phase truncation. This last value
yields a linewidth of 30~mHz over a 100~kHz bandwidth. The corresponding relative frequency noise level is 90~dB
below the requirements of our RF-dressed trapping experiment.

The amplitude noise of the DDS synthesizer is measured to be below $-130$~dB.Hz$^{-1}$ at frequencies above 1~kHz.
Amplitude noise turns out to be the limiting noise present in our RF chain at the present time due to the several dB
noise figures of the attenuators and amplifiers in use in our experiment. They actually increase the RF amplitude noise
to $-115$~dB.Hz$^{-1}$ at 1~kHz, which still remains 25~dB below our experiment requirements.

All these features make our DDS apparatus very well-suited for cold atom experiments. It may be used for
optimized evaporative cooling implementation, with arbitrary frequency ramp profiles including a final ``RF-knife'', in a standard BEC experiment. It
is of particular importance in the case of atom chip experiments~\cite{Folman2002}, where the cooling ramp is usually
done very quickly (1~s typically). It is also ideal for developing new RF-based traps. In particular, the possibility
of configuring the time step, that is the duration of each frequency point, is very attractive for our application. This
feature is all the more interesting in that the memory can be separated in several zones, each being allocated a
chosen number of points and a chosen time step, so we can dilate or compress in time the whole or a part of the ramp,
while keeping the same frequency series. For instance, if we want to accomplish a ramp of $N$ frequency values with a
small time resolution, say 10~$\mu$s, and then keep the final frequency constant over ten seconds, we
have the ability to address a first memory zone with the first $N-1$ points and a 10~$\mu$s time step, and a second
memory zone with the last frequency value and a 10~s time step. A single time step over the whole sequence would have
forced us to sacrifice either the resolution of the ramp or the duration of the final plateau.

The device was successfully implemented in our ultracold atom experiment. First, we set boundary specifications on the
RF source performances for trapping experiments regarding heating rate and lifetime. In the same manner, we set
conditions on maximum phase hops and frequency step magnitude. With our home made device, we observed no heating due
to the RF source during the plateau step. A remaining heating rate of $0.47\mu {\rm K/s}$ was identical to the one
obtained in the magnetic trap and was limited by other noise sources, presumably scattered photons. Despite the very
large number of frequency points, heating along the vertical axis is still present during the loading stage of the
RF-dressed trap. We attribute this heating to non-adiabatic deformation of the confining potential during the loading
process, which we could not make slower due to the other noise sources.

While the actual performances of the device are already very good, a few improvements are still possible. First, the memory
length of the DDS is basically determined by the FPGA in-board memory size. A larger sample size may easily be implemented
by replacing the board, provided that the data transfer rate is improved through the use of an Ethernet port or a USB
port. Second, the sample rate is 1~MHz, limited by the data transfer rate from the FPGA board. Higher FPGA clock
frequency, or smaller clock counts $n_{cycle}$, would allow higher sample rates.
The frequency limit due to the AD-9851 DDS chip itself is 3~MHz. This modification would improve the temporal
resolution by a factor of 3. Finally, the maximum output frequency of 10~MHz may be very easily increased to 20~MHz by replacing the low-pass filter. By clocking the DDS with a 180~MHz clock, one could even reach output frequencies up to 60~MHz. Note that
with a similar DDS chip, the AD 9858, frequency clocks as high as 1 GHz are allowed, which makes possible the synthesis
of RF frequencies up to 330~MHz.

\acknowledgments We are indebted to R.~J.~Butcher for a critical reading of the manuscript. This work was supported by the R\'egion Ile-de-France (contract number E1213) and by the European
Community through the Research Training Network ``FASTNet'' under contract No. HPRN-CT-2002-00304 and Marie Curie
Training Network ``Atom Chips'' under contract No. MRTN-CT-2003-505032. Laboratoire de physique des lasers is UMR 7538
of CNRS and Paris 13 University. The LPL is a member of the Institut Francilien de Recherche sur les Atomes Froids.


\begin{thebibliography}{99}
\bibitem{note1} The factor $B_{\rm RF}/2$ arises from the fact that only half of the power of the linearly polarized RF has an effect on the atoms.
\end{thebibliography}
\end{document}